\documentclass[a4paper]{jpconf}
\usepackage{graphicx}
\begin{document}
\title{Geosynchrotron radio emission from CORSIKA-simulated air showers}

\author{T. Huege, R. Ulrich, R. Engel}

\address{Forschungszentrum Karlsruhe, Inst.\ f\"ur Kernphysik, Postfach 3640, 76021 Karlsruhe, Germany}

\ead{tim.huege@ik.fzk.de}

\begin{abstract}
Simulations of radio emission from extensive air showers we have published so far were performed with a Monte Carlo code using analytical parametrisations to describe the spatial, temporal, energy and angular particle distributions in the air shower. A new version of our simulation code, which replaces these parametrisations with precise, multi-dimensional histograms derived from per-shower CORSIKA simulations, is now available. The new code allows an independent selection between parametrisation and histogram for each of the relevant distributions, enabling us to study the changes arising from using a more realistic air shower model in detail. We describe the new simulation strategy, present some initial results and discuss the new possibilities. \end{abstract}

\section{Introduction}

In the past few years, we have modelled geosynchrotron radio emission from extensive air showers with both analytic calculations \cite{HuegeFalcke2003} and extensive Monte Carlo simulations \cite{HuegeFalcke2005a}, the results of which have been presented and summarised in a parametrisation formula in \cite{HuegeFalcke2005b}. These Monte Carlo simulations were based on an air shower model using analytical parametrisations (see \cite{HuegeFalcke2003}) to describe the spatial, temporal, energy and angular distributions of the air shower particles. The new Monte Carlo code presented here replaces these analytical parametrisations with multi-dimensional histograms generated for each individual shower by the air shower simulation code CORSIKA \cite{Heck1998}, allowing a much more precise calculation of the radio signal.

In the following, we present the new simulation strategy before discussing a few examples of the effects arising as a consequence of the transition to the realistic, histogrammed particle distributions and the elimination of the track length parameter that was present in the earlier simulations. Afterwards, we discuss some examples of how the new code can be used to explain the shape of the emitted radio pulses, followed by our conclusions.

\section{CORSIKA-based Monte Carlo simulations}

In our new Monte Carlo code, REAS V2, the process of simulating radio emission from extensive air showers is separated into two steps. First, one simulates the air shower as usual with CORSIKA, using the desired interaction models and simulation parameters. A tailor-made interface code collects the particle histogram information during the air shower simulation and outputs it to disk in a compact data file. The particle information is sampled in (currently) 50 slices distributed over the shower evolution equidistantly in (slant) atmospheric depth. In each of these slices, four three-dimensional histograms are filled: for electrons and positrons separately one histogram of particle lateral distance vs.\ particle arrival time vs.\ particle energy and one histogram of particle momentum angle to the shower axis vs.\ particle momentum angle to the radial direction vs.\ particle energy. Separating the information into two histograms (as opposed to one higher-dimensional histogram) is a reasonable approximation, as the specific choice for separating the parameters is motivated by air shower physics. In a second step, the histograms and a detailed longitudinal shower profile are used to create particles which fulfill these distributions and calculate the radio emission from their deflection in the earth's magnetic field. The well-understood and thoroughly tested radio emission calculation is unchanged from our earlier code (see \cite{HuegeFalcke2005a}).

This two-step approach has a number of advantages. In particular, it allows us to make a very gradual transition from fully parametrised air showers to fully histogrammed air showers by switching on the histogrammed distributions individually one after the other and evaluating the changes arising in the radio signal. The fact that one does not have to re-run the complete air shower simulation each time one wants to calculate the radio emission in a different configuration is also very helpful in keeping the computation time low.

\section{Effects of the realistic distributions}

To illustrate the changes between the earlier (REAS V1, cf.\ \cite{HuegeFalcke2005a,HuegeFalcke2005b}) and new simulations, we look at a prototypical vertical $10^{17}$~eV proton-induced air shower. Pulses shown here are for an observer at 75~m and an observer at 525~m to the north from the shower centre (in the latter case amplified by a factor of 20), illustrating two different regimes. Effects concerning other azimuthal directions are beyond the scope of this article and will be discussed elsewhere. We show three particularly interesting examples of effects arising when changing from the parametrised distributions to the histogrammed distributions. In the following, we compare the radio emission from air showers with identical particle distributions (the longitudinal air shower evolution is histogrammed, all other distributions are parametrised), except for one specific distribution that we switch from parametrised to histogrammed to assess the effect on the radio signal.

\subsection{Particle energies}

\begin{figure}[htb]
\vspace{-1pc}
\begin{minipage}{18pc}
\includegraphics[angle=270,width=18pc]{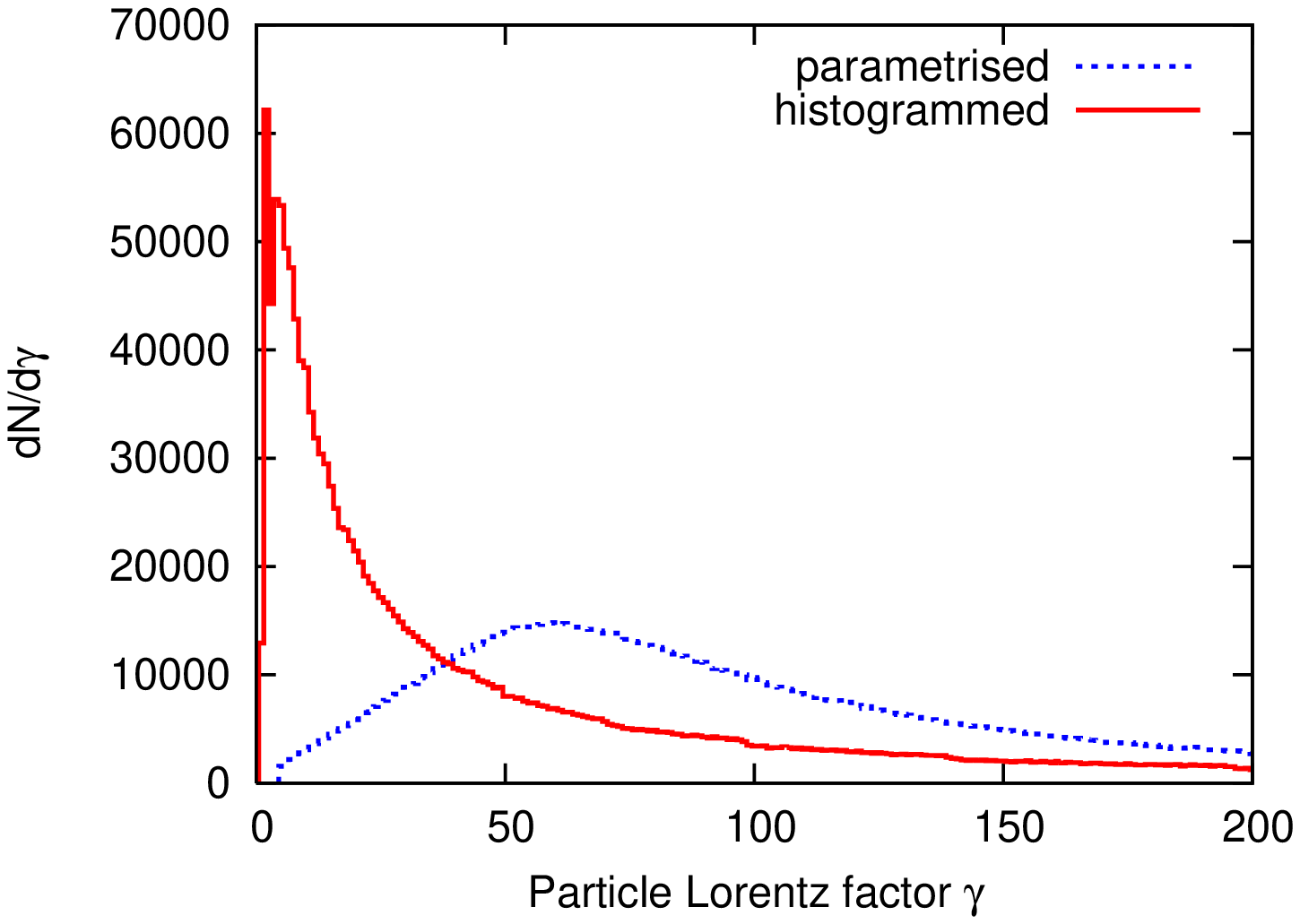}
\caption{\label{energyhisto}Distribution of electron energies in parametrised and histogrammed versions.}
\end{minipage}\hspace{2pc}%
\begin{minipage}{18pc}
\includegraphics[angle=270,width=18pc]{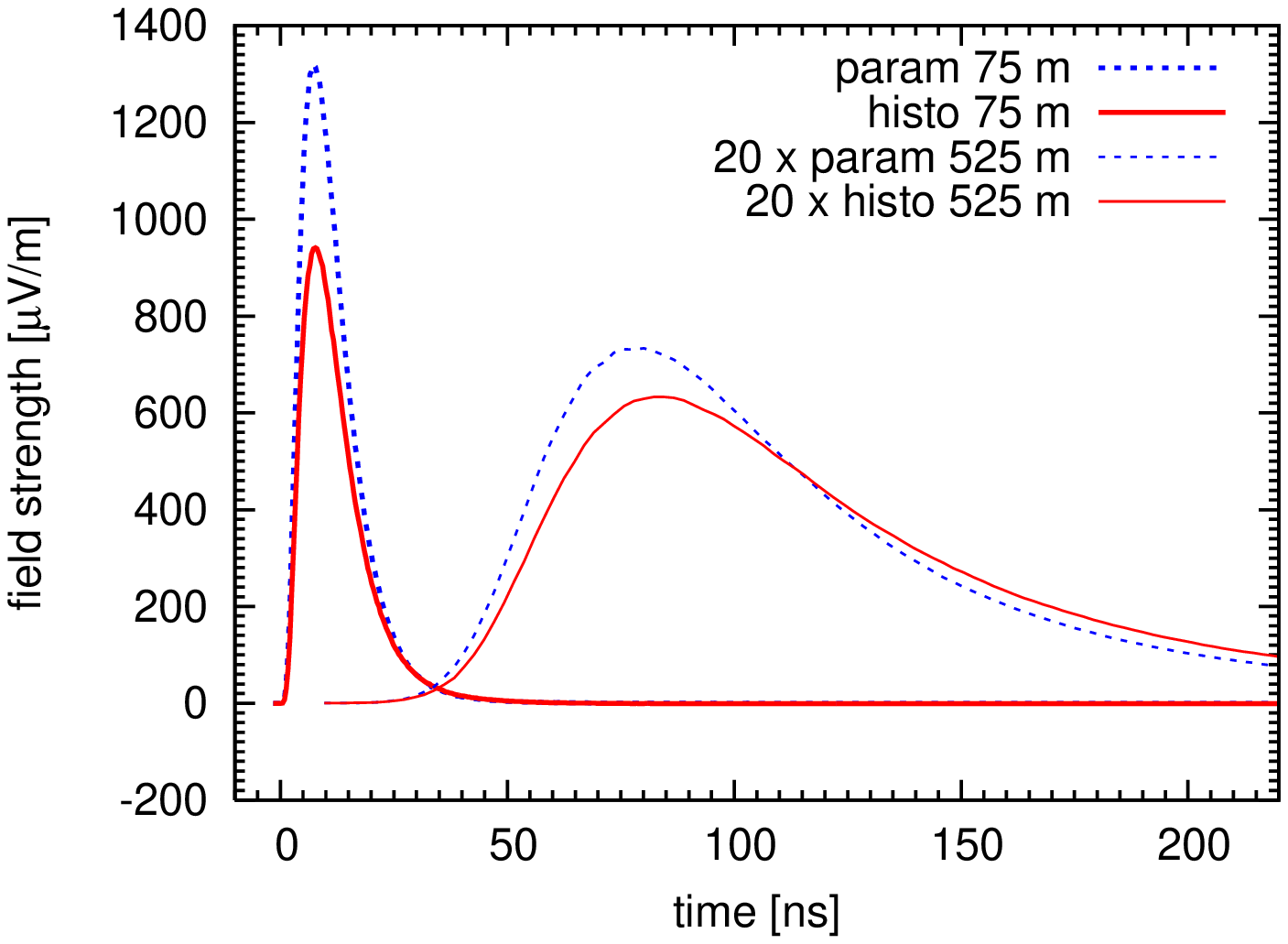}
\caption{\label{energypulses}Radio pulses for parametrised and histogrammed particle energy distributions.}
\end{minipage} 
\end{figure}

\noindent Figure \ref{energyhisto} shows the parametrised particle energy distribution used in the earlier simulations \cite{HuegeFalcke2005a,HuegeFalcke2005b} in comparison with the histogrammed electron energy distribution as obtained from CORSIKA. (We only show electron distributions here, the positron distributions are similar. If not stated otherwise, the histograms are averaged over shower evolution and spatial coordinates.) The histogrammed energy distribution deviates significantly from the earlier parametrisation. In particular, lower energies are much more pronounced and there is no peak in the distribution at medium Lorentz factors. Figure \ref{energypulses} shows the corresponding radio pulses. The changes are rather slight. The relative up-weighting of low energies, and, correspondingly, down-weighting of higher energies is mainly visible in a reduced pulse height in the shower centre, where high-energy particles with narrow beaming cones contribute strongly (cf.\ Sec.\ \ref{sec:pulseshapes}).

\subsection{Particle arrival times}

\begin{figure}[htb]
\vspace{-1pc}
\begin{minipage}{18pc}
\includegraphics[angle=270,width=18pc]{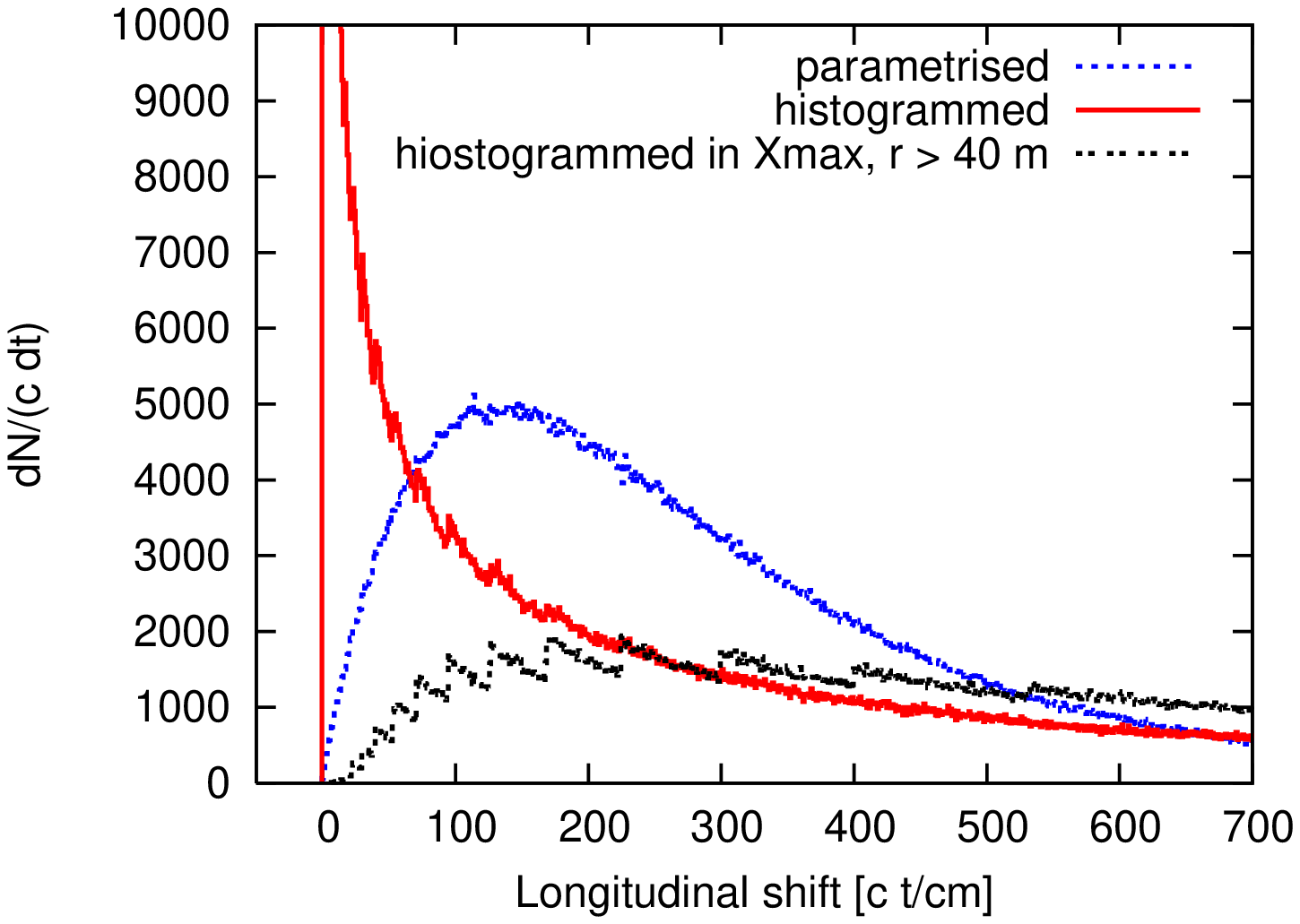}
\caption{\label{timehisto}Parametrised and histogrammed electron arrival time distributions.}
\end{minipage}\hspace{2pc}%
\begin{minipage}{18pc}
\includegraphics[angle=270,width=18pc]{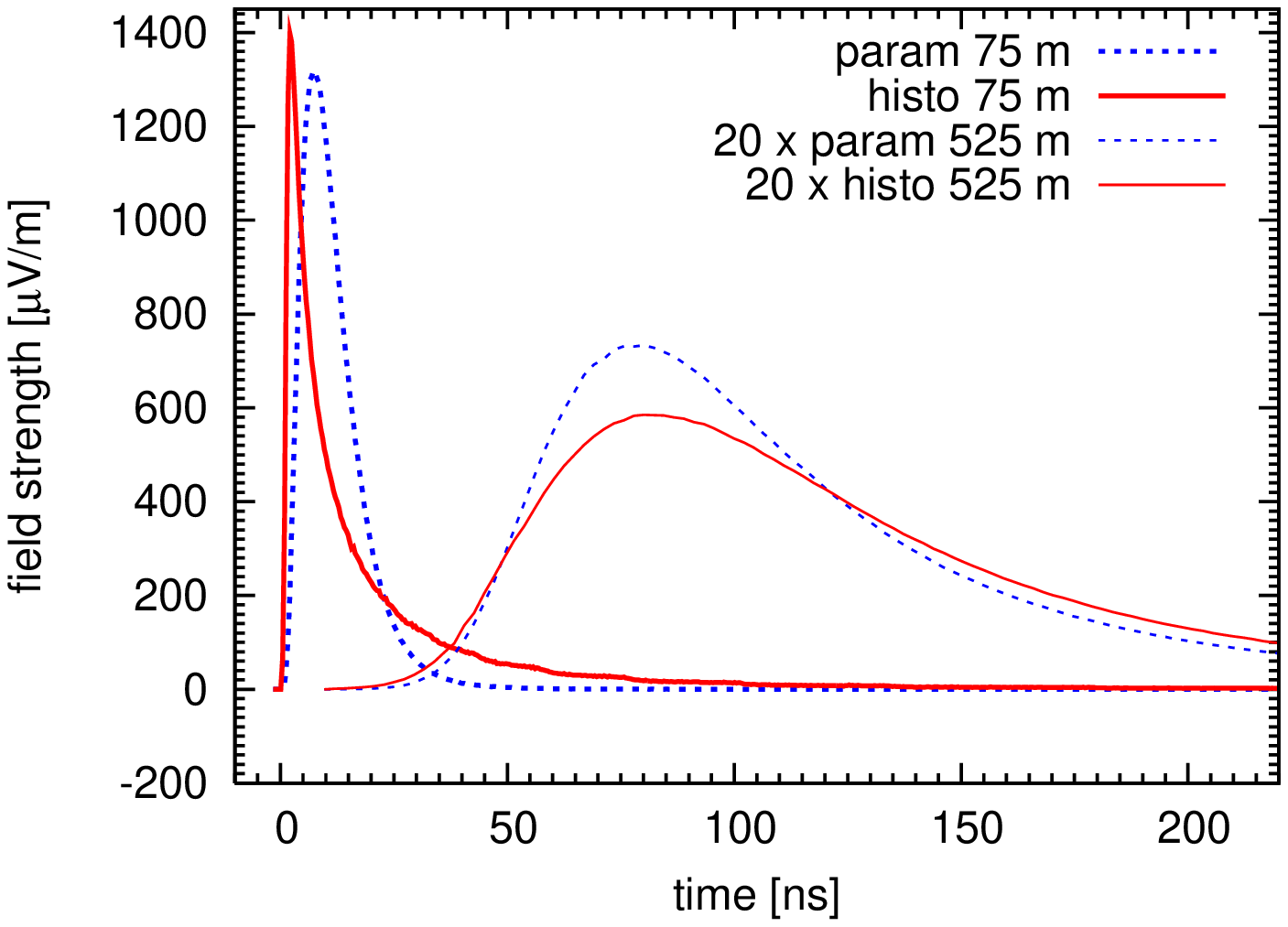}
\caption{\label{timepulses}Radio pulses for parametrised and histogrammed particle arrival times.}
\end{minipage} 
\end{figure}

\noindent The arrival time distribution has strong influence on the shape (and thus frequency spectrum) of the radio pulses, for a discussion see \cite{HuegeFalcke2003}. Figure \ref{timehisto} shows the parametrisation of the electron arrival times used earlier in comparison with the distribution obtained from CORSIKA. The CORSIKA-based arrival time distribution is much narrower. This was to be expected, as the parametrisation \cite{HuegeFalcke2003} was based on experimental measurements of the distributions, which are only available at relatively large distances from the shower core, where the distributions are systematically wider. Also, the parametrised distribution was derived from measurements of all charged particles, including muons, which could additionally widen the distribution. In the shower centre, the CORSIKA-based distribution is extremely spiked and qualitatively different from the parametrisation. At larger radial distances, the CORSIKA-derived distribution becomes qualitatively similar to the parametrisation (and thus measurements).

Please note that the ``zig-zag'' structure present in the histogrammed distributions is an artifact of the adopted logarithmic binning. As the logarithmic bin boundaries have no regular structure on a linear scale, this is unproblematic for the calculation of the radio signal. In fact, in most cases this is actually favourable and helps to recreate the underlying distribution well.

The differences in particle arrival time directly affect the shapes of the radio pulses as plotted in Fig.\ \ref{timepulses}. In particular, close to the shower centre, where geometrical time delays are of no importance, the pulse shape directly reflects the particle arrival time distribution and thus becomes much narrower in the histogrammed version. (Atmospheric refraction which is expected to slightly smear out the pulses is not taken into account here.)

\subsection{Particle momentum angles to shower axis}

\begin{figure}[htb]
\begin{minipage}{18pc}
\includegraphics[angle=270,width=18pc]{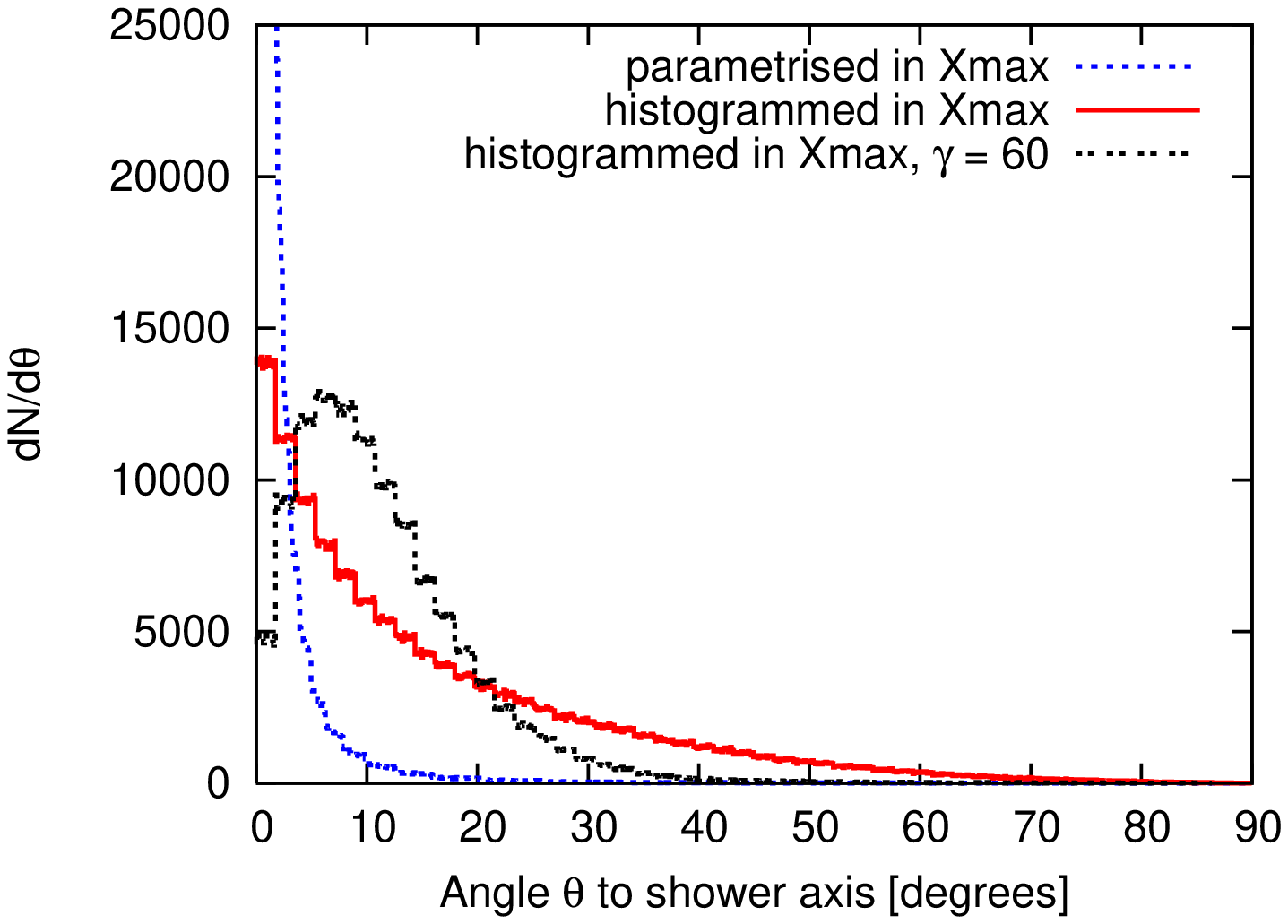}
\caption{\label{anglehisto}Distribution of electron momentum angles to the shower axis.}
\end{minipage}\hspace{2pc}%
\begin{minipage}{18pc}
\includegraphics[angle=270,width=18pc]{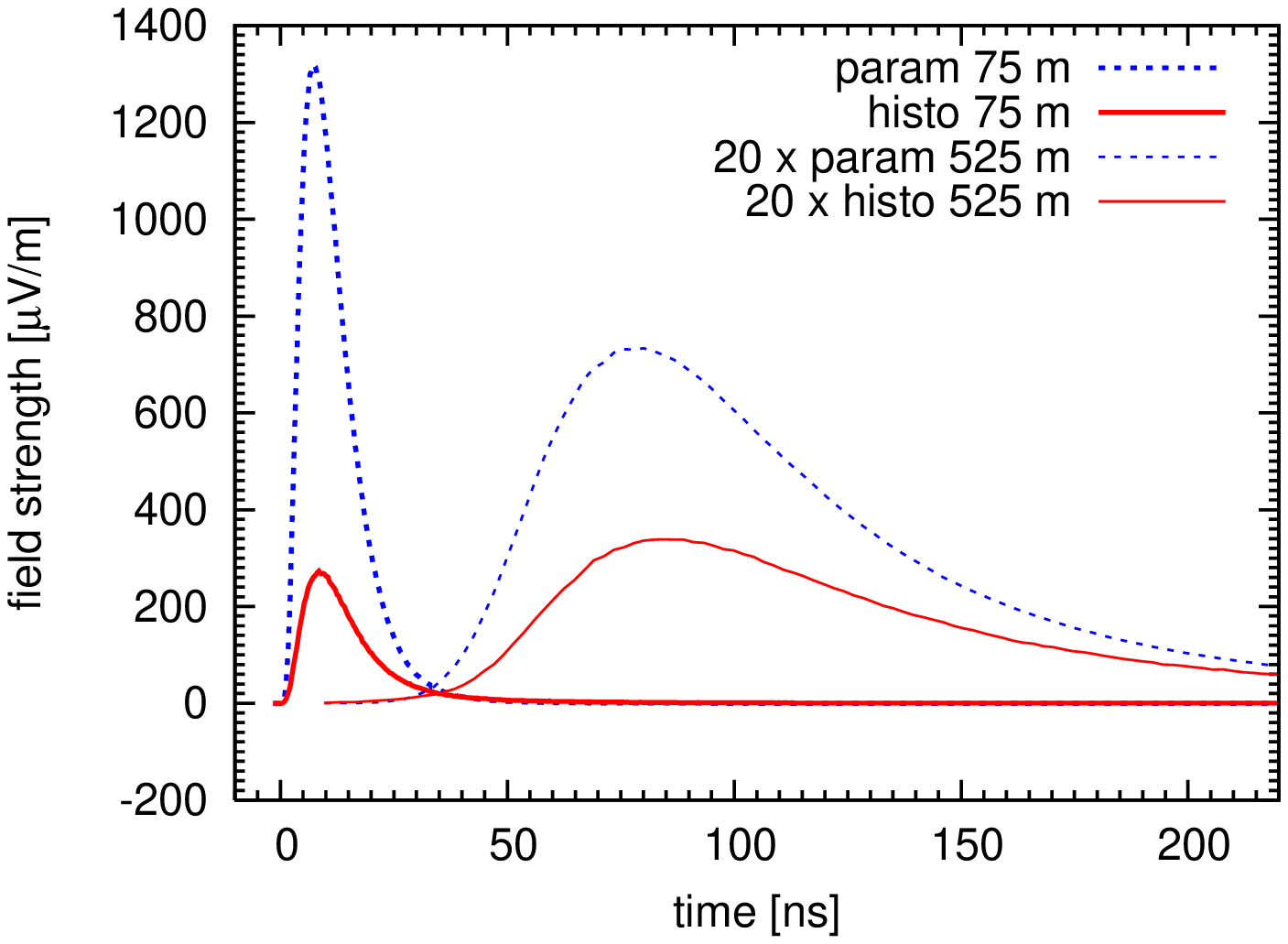}
\caption{\label{anglepulses}Radio pulses for parametrised and histogrammed particle momentum angles.}
\end{minipage} 
\end{figure}

\noindent Figure \ref{anglehisto} shows the angular distribution of electron momenta with regard to the shower axis. In the parametrised version, the particle momenta were assumed to be very collimated and directly correlated to the lateral distance form the shower axis. The histogrammed distribution is much wider and strongly dependent on the particle energy as illustrated by a comparison of the overall distribution with that of electrons with a Lorentz factor of 60. Correspondingly, and expectedly, the effect on the radio pulses is strong. Since the emission of the highly relativistic particles is beamed into a narrow cone along the particle momentum direction, a significant fraction of the radio signal is now emitted at large angles to the shower axis and therefore does not contribute to the radio pulses measured at the distances from the shower centre that are of relevance to detection with ground-based arrays. The pulses are therefore damped significantly.

\section{Eliminating the track length parameter}

\begin{figure}[htb]
\vspace{-0.8pc}
\begin{minipage}{18pc}
\includegraphics[angle=270,width=18pc]{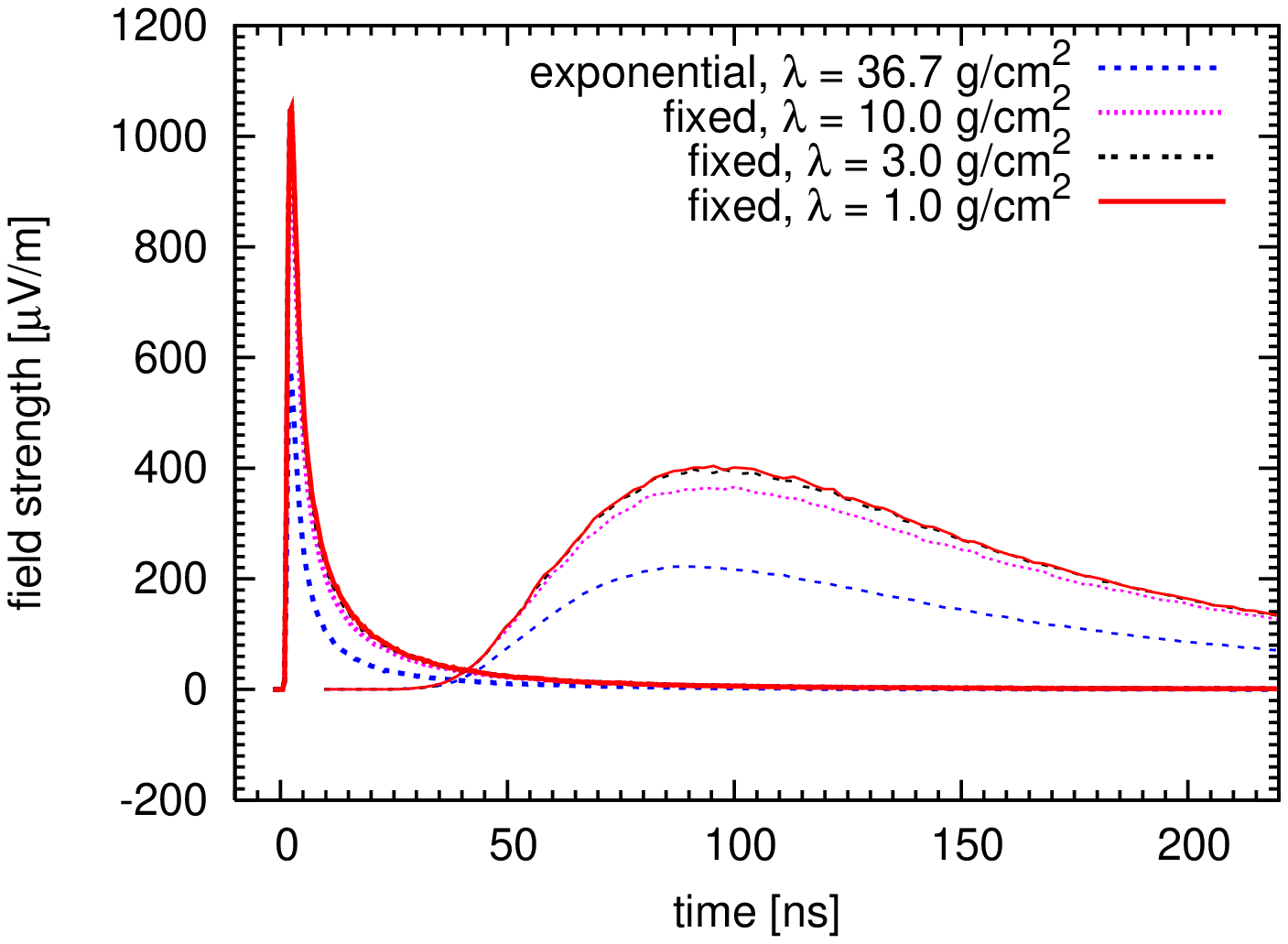}
\caption{\label{tracklengthpulses}Radio signal convergence going from 36.7~g~cm$^{-2}$, exponentially distributed, to 1.0~g~cm$^{-2}$, fixed track lengths.}
\end{minipage}\hspace{2pc}%
\begin{minipage}{18pc}
\includegraphics[angle=270,width=18pc]{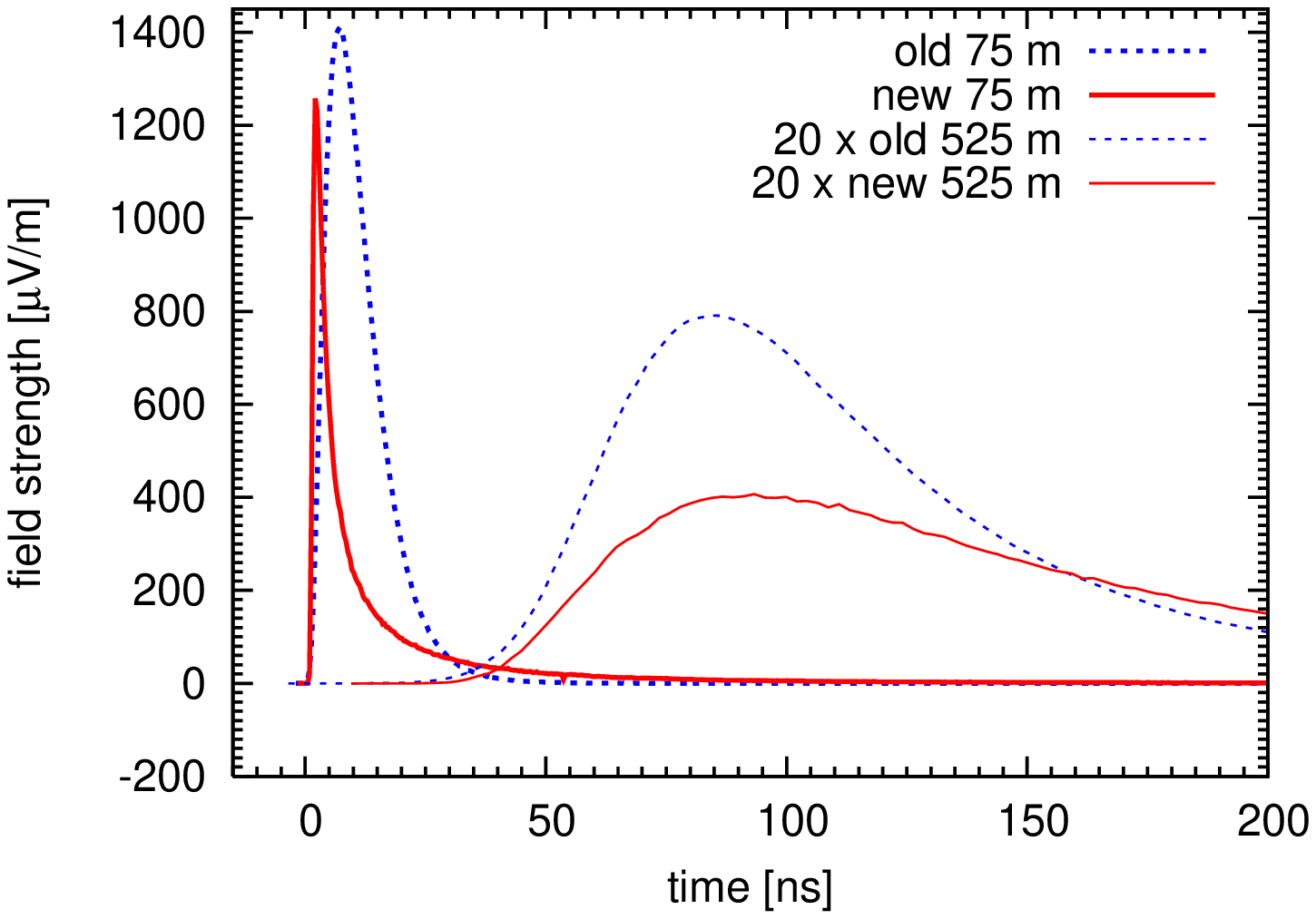}
\caption{\label{overallpulses}Overall change in the radio pulses from fully parametrised (``old'') to fully histogrammed (``new'') air showers.}
\end{minipage} 
\end{figure}

\noindent The earlier simulations contained a free parameter, the so-called ``track length parameter'' $\lambda$, which is neither easily accessible from experiments nor from simulations. A large value leads to a simulation with few long tracks, whereas a small value corresponds to a simulation with many short tracks. Now that very detailed particle distributions are available, the parameter can be eliminated by simulating a high number of very short tracks. Long tracks are then effectively described by multiple short, but representative, segments. This ensures that the particles always represent the local distributions well throughout their complete trajectory. If this picture is correct, the radio signal should converge when one goes from exponentially distributed long track lengths (as in the old simulations) to shorter and shorter fixed track lengths, where the result becomes independent of $\lambda$. This is exactly what is visible in Fig.\ \ref{tracklengthpulses}.

\section{Overall result}

\begin{figure}[htb]
\vspace{-1pc}
\begin{minipage}{18pc}
\includegraphics[angle=270,width=18pc]{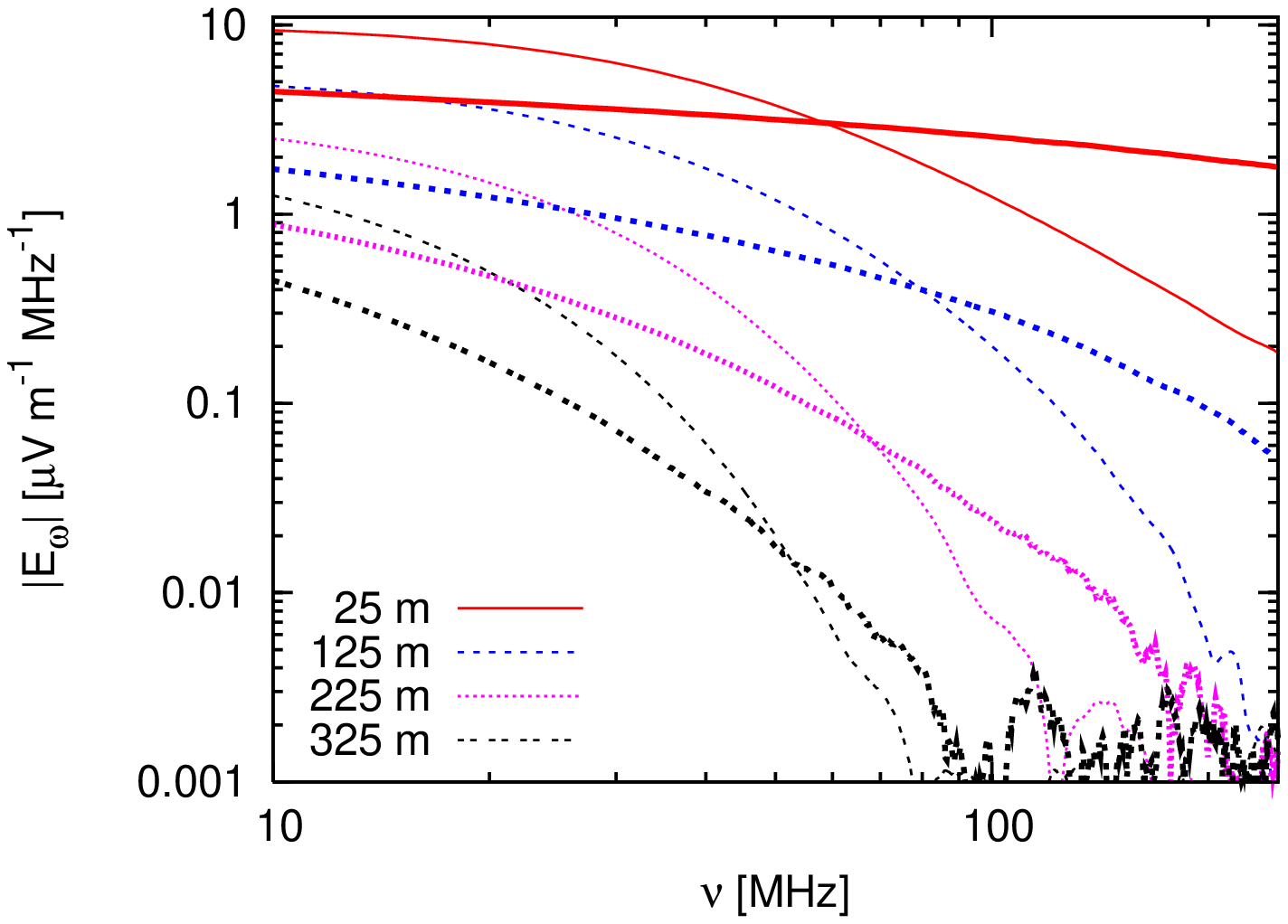}
\caption{\label{overallspectra}Overall change in the radio signal frequency spectra from a fully parametrised (thin lines) to a fully histogrammed (thick lines) air shower at various distances to the north from the shower centre.}
\end{minipage}\hspace{2pc}%
\begin{minipage}{18pc}
\includegraphics[angle=270,width=18pc]{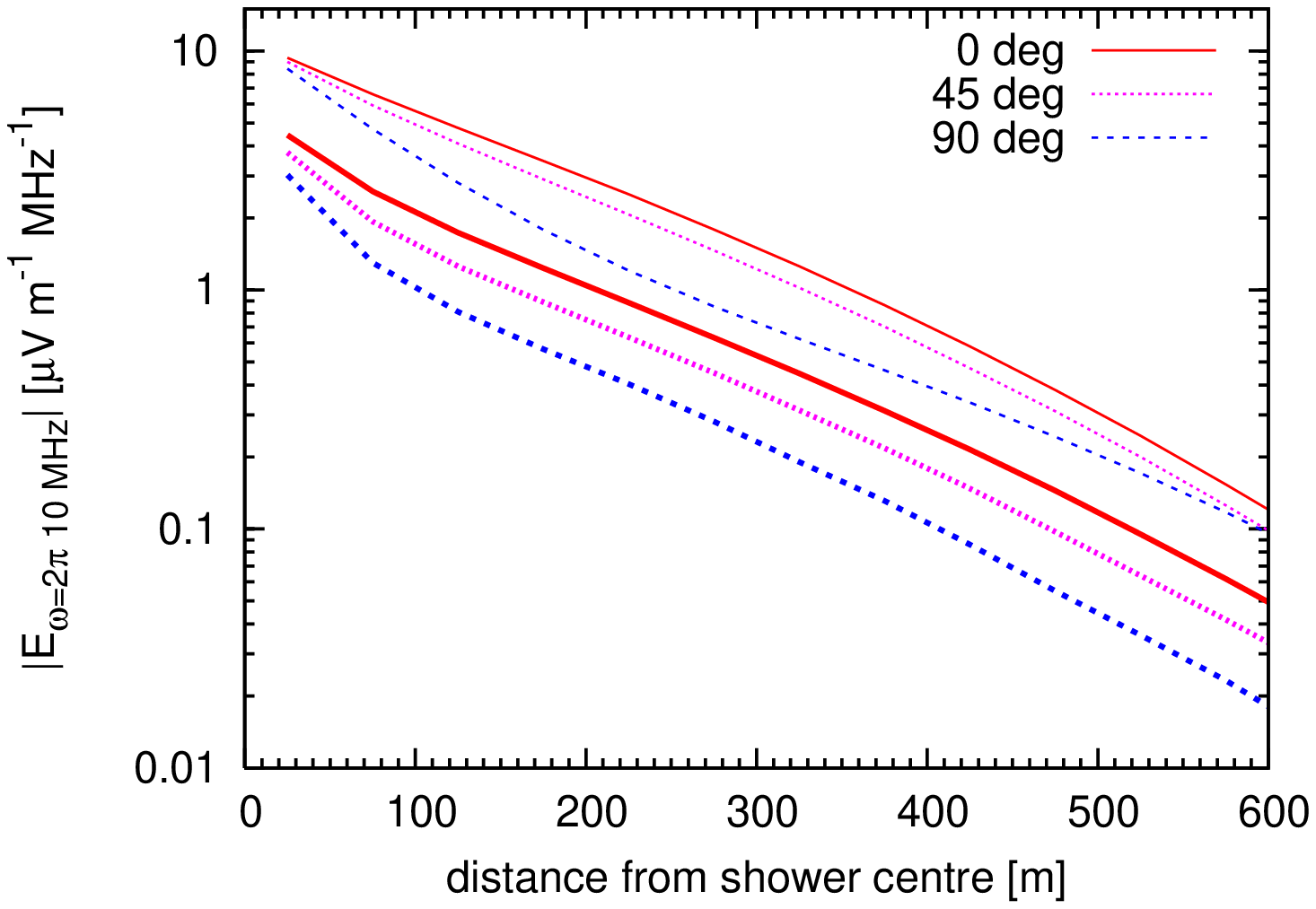}
\caption{\label{radial}Spectral field strength at 10~MHz as a function of radial distance to the shower centre for different azimuthal directions (0$^{\circ}$ is north, 90$^{\circ}$ is west) compared between old (thin lines) and new (thick lines) simulations.}
\end{minipage} 
\end{figure}

\noindent In Fig.\ \ref{overallpulses} we compare our ``old'' simulations (fully parametrised shower as in REAS V1, $\lambda$ set to 36.7~g~cm$^{-2}$ with exponential distribution, cf.\ \cite{HuegeFalcke2005b}) with the ``new'' simulations (REAS V2, fully histogrammed shower, $\lambda$ set to 1.0~g~cm$^{-2}$ fixed). The changes are significant, but not dramatic. There is a visible damping of the pulses, and the pulses close to the shower centre are much narrower, a consequence of the histogrammed arrival time distribution. Looking at the results in the frequency domain (Fig.\ \ref{overallspectra}), the narrower pulses close to the shower centre are reflected by a flatter frequency dependence extending to higher frequencies. This is favourable for an experimental detection of radio signals, as there is a natural limit at low frequencies given by atmospheric and ionospheric noise. Atmospheric refraction could, however, slightly smear out the pulses in the centre region and thus steepen the spectra once again. The lateral dependence of the radio emission is plotted in Fig.\ \ref{radial}. The lateral decrease of the 10~MHz spectral field strength is very similar in both the new and old simulations. However, there is a more pronounced north-south to east-west asymmetry visible in the new simulations.

\section{Pulse shape analysis} \label{sec:pulseshapes}

\begin{figure}[htb]
\begin{minipage}{18pc}
\includegraphics[angle=270,width=18pc]{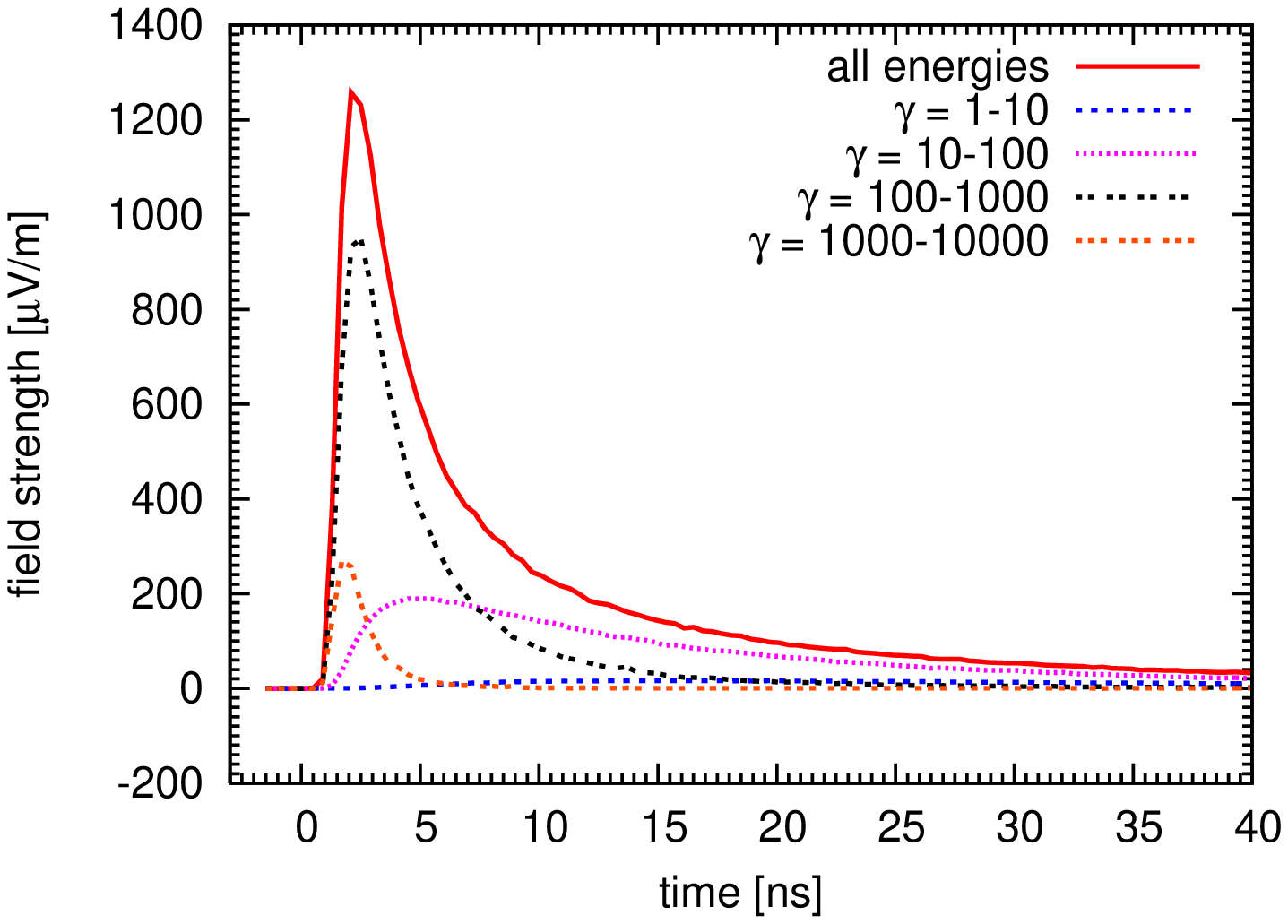}
\caption{\label{energyregimes}Contribution of different particle energy regimes to the radio pulse at 75~m north from the shower centre.}
\end{minipage}\hspace{2pc}%
\begin{minipage}{18pc}
\includegraphics[angle=270,width=18pc]{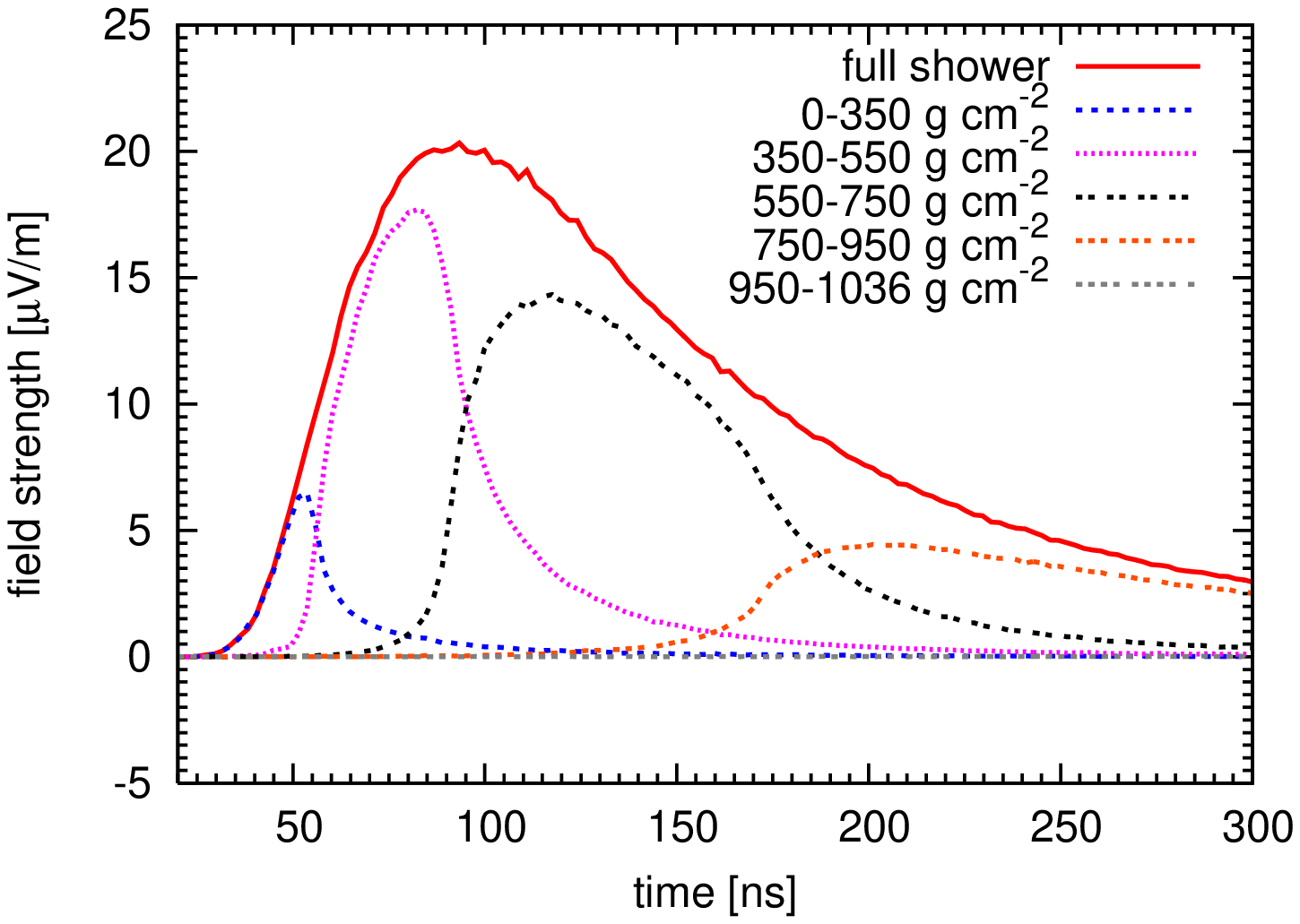}
\caption{\label{depthregimes}Contribution of different shower evolution stages to the radio pulse at 525~m north from the shower centre.}
\end{minipage} 
\end{figure}

Now that detailed information about the particle distributions is available throughout the full air shower evolution, it is possible to look deeper into the structure of the radio time pulses and their relation to shower particle distributions. Figure \ref{energyregimes} shows the contributions of different particle energy regions to the overall radio pulse close to the shower centre. It is evident that high-energy particles with Lorentz factors between 100 and 1000 dominate the emission, although low-energy particles by far outnumber them (cf.\ Fig.\ \ref{energyhisto}). This is clear because higher energy particles radiate more efficiently than low-energy particles, but only into a narrow beaming-cone encompassing only the centre region. Consequently, the pulse at 525~m (not shown here) is dominated by emission from lower energy particles with Lorentz factors between 10 and 100.

Another option is to look at the contribution of different stages of the air shower evolution to the radio signal. Because of geometrical time-delays between radiation originating from different parts of the air shower, this is especially interesting at larger distances from the shower centre. Figure \ref{depthregimes} shows the radio pulse at 525~m to the north from the shower centre. One can clearly differentiate the contributions from the different air shower stages. The emission is dominated by the shower stage around the shower maximum (which here resides at $\sim640$~g~cm$^{-2}$) and the stage shortly before the shower maximum. This is somewhat surprising as the particle numbers before and after the shower maximum are similar but the particles after the shower maximum are closer to the observer and thus could have been expected to contribute more strongly. The explanation is that the particles in the stage before the shower maximum traverse a much thinner medium and thus propagate for longer (geometrical) distances, during which they can emit radio signals. In general, a thinner medium leads to higher geosynchrotron radio emission --- if the shower is able to evolve to high particle multiplicities. The contributions to the radio pulse by different stages of the air shower evolution show that, in principle, information on the air shower evolution (such as the position of the shower maximum, which in turn is related to the mass of the primary particle) is encoded in the radio signal. This could possibly be exploited in radio measurements of cosmic ray air showers. The most favourable regime for such a measurement would be at distances of more than a few hundred metres from the shower centre.

Other analyses like these can be readily performed with REAS V2. Also, the influence of the primary particle mass and shower to shower fluctuations can now be easily evaluated. These and other aspects such as azimuthal asymmetries and polarisation characteristics will be covered in much more detail in an upcoming publication.

\section{Conclusions}

We have developed a next-generation Monte Carlo code for the calculation of geosynchrotron radio emission from CORSIKA-simulated cosmic ray air showers. The transition from our earlier code to this new version is gradual, allowing to understand the changes in the radio pulses in detail. A last free parameter remaining in our earlier code has been eliminated in a natural, self-consistent way. The adopted two-step approach proves to be very powerful in many situations. 

Although there are significant differences in the particle distributions calculated by CORSIKA in comparison to the parametrisations used earlier, the changes of the resulting radio pulses are not dramatic. Along the north-south observer axis, a damping of about a factor of two and a somewhat flatter frequency distribution in the shower centre arise.


\section*{References}

\begin{thebibliography}{9}
\bibitem{HuegeFalcke2003} Huege T and Falcke H 2003 {\it Astronomy \& Astrophysics} {\bf 412} 19
\bibitem{HuegeFalcke2005a} Huege T and Falcke H 2005 {\it Astronomy \& Astrophysics} {\bf 430} 779
\bibitem{HuegeFalcke2005b} Huege T and Falcke H 2005 {\it Astropart. Phys.} {\bf 24} 116
\bibitem{Heck1998} Heck D et al. 1998 {\it Forschungszentrum Karlsruhe Report FZKA} {\bf 6019} 
\end{thebibliography}
\end{document}